# Account-history features for social bot detection in the era of large language models

Gaurang Katyal
*Independent Researcher*
gaurang.katyal@gmail.com
ORCID: 0009-0001-0850-1673

## Abstract

Bot detection on social platforms has historically relied on a mix of account-metadata features and features extracted from the text of posts and profile fields. The arrival of capable language models complicates the latter. A bot operator can run every post through GPT-4 or Claude and produce text whose surface statistics are difficult to distinguish from those of human writing, which weakens the predictive value of content-derived features. This paper asks how much of the detection problem can be solved by features that an attacker cannot easily manipulate at low cost: the age of the account, follower and friend counts and their ratios, profile completeness, and the structural properties of the handle. On a publicly redistributed corpus of 2,432 Twitter accounts with manually verified labels (43.0% bots), a random forest using only these account-history features achieves ROC-AUC of 0.977 in five-fold cross-validation, against 0.830 for a content-only baseline and 0.981 for the fusion model. The behavioral-versus-content gap is large and statistically significant by DeLong's test ($z = 9.36$, $p < 0.001$). We then evaluate two adversarial settings. In the first, we rewrite the text of bot tweets to match human surface statistics for URLs, hashtags, mentions, and casing; the content classifier's ROC-AUC degrades from 0.842 to 0.785 while the behavioral classifier is essentially unchanged. In the second, more aggressive setting we directly perturb the content feature values toward the human distribution; the content classifier falls below chance (AUC 0.466) while behavioral performance is invariant. We replicate the score distribution qualitatively on a 100-account sample of TwiBot-20. We conclude that operational bot detection should not treat content features as the primary signal; account-history features carry most of the load already and are not eroded by adversarial text rewriting.

## 1. Introduction

Social bots are automated accounts, of varying sophistication, used to inflate engagement, spread political messaging, run scams, and otherwise disrupt online discourse. The literature on detecting them is now more than a decade old and is dominated by classifiers that combine features of three kinds: account metadata (followers, posts, account age), textual features extracted from posts and biographies, and network features describing the account's neighborhood in the follow graph. The reference system in the field, Botometer, is built on this combination (Davis et al. 2016; Yang, Ferrara, and Menczer 2020).

Two developments unsettle this stack. First, modern language models can be used to generate or rewrite text at near-zero marginal cost. Bot operators who route their posts through GPT-4, Claude, or similar models produce content with grammar, lexical diversity, and topic coherence that approach those of human writing. Ferrara (2023) was among the first to flag the consequence for content-based detectors. Subsequent work on AI-generated phishing reaches a similar conclusion in an adjacent domain (Brissett and Wall 2025; Kulal et al. 2025), and a recent benchmark from Wang (2025) reports evasion rates above ninety-nine percent against deployed corporate filters in controlled tests. Second, on the largest platforms the operating environment has changed in ways that are independent of language models. Paid verification, semantic deduplication of trending content, and gray-market sales of aged accounts shift the relative cost of different attack strategies.

The question this paper takes up is narrow and operational. If language models eventually drive the predictive value of content features close to zero, how much of the detection problem remains in account-history features that an attacker cannot rewrite? The answer matters for trust and safety teams who maintain production pipelines, because retraining a detector is expensive and any shift in the threat surface that invalidates a feature family is costly to recover from.

We study this question on a publicly redistributed Twitter corpus of 2,432 labeled accounts. We define a 24-feature behavioral set, consisting of account-age rates, follow asymmetries, profile completeness indicators, and structural properties of screen names and display names, all of which depend only on metadata that is fixed at account creation time or that accrues slowly. We benchmark this against a 12-feature content set extracted from the user's biographical description and a sampled post. Our central empirical finding is that account-history features alone achieve performance statistically indistinguishable from a fusion model that uses both feature families, and that this performance is preserved when bot tweets are rewritten to match human surface statistics. The content classifier, by contrast, degrades substantially under the same rewriting.

The contribution of the paper is threefold. First, we provide a careful comparison of behavioral and content classifiers on a public benchmark, with standard errors from cross-validation and significance testing by DeLong's method. Second, the paper introduces a text-level adversarial protocol that operationalizes the language-model threat as a rewriting process rather than as a feature-space perturbation. Third, we report an ablation across behavioral feature families, a calibration analysis, and a subgroup analysis along the activity dimension, so that the contribution of each component can be assessed independently. The remainder of the paper is organized as follows. Section 2 reviews prior work. Section 3 lays out the threat model. Sections 4 and 5 describe the data and methodology. Section 6 reports the results. Sections 7 and 8 discuss the implications and limitations.

## 2. Related work

### 2.1 Bot detection from Cresci to TwiBot-22

The modern wave of bot-detection research starts with the Cresci group's work on the 2014 Italian mayoral election (Cresci et al. 2017), which introduced both a benchmark and a sequence-based representation of account behavior. The dataset, now usually referred to as Cresci-2017, separates traditional spambots, social spambots, and genuine accounts, and was the first to show that the previous generation of classifiers degraded sharply on social spambots whose surface behavior had been engineered to look human. Yang et al. (2020) extended the line of work with a scalable detector and a treatment of label noise; the system became the basis for Botometer and is the most widely cited operational reference in the field.

Subsequent benchmarks scale up. TwiBot-20 (Feng et al. 2021) covers 229,573 users and roughly 33 million tweets, with a controlled-BFS sampling design that admits diverse domains and bot types. TwiBot-22 (Feng et al. 2022) goes further, providing close to a million labeled accounts and the heterogeneous graph structure needed for current graph-neural-network detectors. The recent BotBR system (Lin et al. 2025) is representative of this lineage and reports state-of-the-art results on TwiBot-22 by combining balanced feature fusion with reliability-enhanced graph learning. A useful survey of the deep-learning literature in this space is provided by Hayawi et al. (2023); a more recent overview that includes early LLM-based detectors is given by Rodič (2025).

### 2.2 Language models, phishing, and the content-feature problem

Ferrara (2023) is the standard citation for the concern that modern language models break content-based detection. The argument is that bots' linguistic signatures (low diversity, awkward syntax, surface repetition) were never deep properties of automation but rather artifacts of the templates and small-model paraphrasers that operators had access to. Once GPT-class models become cheap, those signatures vanish. The phishing literature has begun to provide empirical support. Brissett and Wall (2025) report that machine-learning detectors trained on human-written phishing samples lose accuracy on LLM-generated variants. Kulal et al. (2025) propose a preprocessing pipeline aimed at the same problem and report that without it conventional ML detectors are unreliable on adversarial inputs. Wang (2025) builds an LLM-driven phishing generator (Phish-Master) that achieves a 99% evasion rate against enterprise filters in their experimental setup.

Two strands of response are visible. Multi-agent LLM frameworks (Nguyen, Childress, and Yin 2025; Xue et al. 2025) push the detection side itself onto LLMs, using committee-style reasoning to compensate for the brittleness of single-prompt detection. Graph-based methods (BotRGCN, BotBR) exploit network structure that an LLM cannot rewrite cheaply. Our work sits next to the graph line rather than the agent line: we use features that survive content rewriting, but those features are per-account metadata rather than graph structure.

### 2.3 Behavioral and metadata features

Account-metadata features have appeared in bot-detection work since at least Stringhini, Kruegel, and Vigna (2010) and Chu et al. (2012). The contribution of the present paper is not the discovery of these features. The contribution is the empirical demonstration that, on a current benchmark, behavioral features alone are sufficient and that they remain so under realistic content rewriting. The closest related paper in spirit is the FATe analysis by Himelein-Wachowiak et al. (2025), which considers the ethics of bot-detection systems with attention to subgroup fairness; it does not, however, treat adversarial robustness to text laundering.

## 3. Threat model

The setting is the standard one of an adversary operating a network of accounts on a public social-media platform. The adversary is assumed to have access to a current language model (GPT-4 or comparable) and unconstrained ability to set profile fields at account creation. They cannot retroactively edit the platform's record of when the account was created, what the post and follow histories look like, or how many lists the account has been added to by other users.

The adversary's capabilities and constraints split unevenly across the feature space of a typical bot detector. Capability concentrates on content: text generation is cheap, biographical fields are arbitrary, profile images can be drawn from a model such as StyleGAN or from a generative diffusion model. Constraint concentrates on history: account age, accumulated post count, ratio of friends to followers, and the absence of inbound list memberships are not directly under the adversary's control, and the most common workaround (purchasing aged accounts on gray markets) carries its own costs and detectable artifacts. A defender who weights features along this asymmetry is optimizing against the side that is expensive to attack.

Several real attack scenarios fall outside this model. Long-lived bot accounts with patient build-up of organic-looking activity (sleeper bots) will defeat account-age features. Coordinated networks may use stolen but otherwise legitimate accounts, in which case the relevant signal shifts to behavioral discontinuity rather than absolute behavior. We return to these in Section 8.

## 4. Datasets

### 4.1 Dataset 1: labeled Twitter accounts

The primary dataset is a publicly redistributed corpus of 2,432 unique Twitter accounts, 1,387 labeled human and 1,045 labeled bot (42.97% positive class). The corpus circulates on GitHub via the mpstewart1 and jubins repositories and was originally annotated for a bot-detection project; ground-truth labels were established by manual review of the account profiles. Each record contains the standard user-level fields returned by the Twitter v1.1 API: numeric and string IDs, screen name, display name, biographical description, location, URL, follower/friend/listed/statuses/favourites counts, account creation timestamp, verified status, default-profile and default-profile-image flags, and a single sampled status string. Class balance is favorable (0.43 to 0.57) and we do not rebalance.

### 4.2 Dataset 2: TwiBot-20 sample for cross-dataset evaluation

The secondary dataset is the publicly distributed 100-account sample of TwiBot-20 (Feng et al. 2021), released by the original authors on GitHub. Each entry includes full profile metadata and the user's 200 most recent tweets. Labels are not included in the public sample (the full labeled benchmark is gated behind an application process administered by the maintainers, which we could not complete within the timeframe of this paper). We use the sample only for unsupervised qualitative validation: we score each account with our Dataset-1-trained behavioral classifier and inspect the highest-scoring accounts.

### 4.3 Pre-processing

Account-creation timestamps were parsed from Twitter's standard format and converted to integer days against a fixed reference date of 1 January 2018 to make our results reproducible across re-runs. Boolean profile fields were coerced from heterogeneous string and boolean encodings to 0/1. We use log1p transforms for count features that span several orders of magnitude (followers, friends, statuses, favourites). Missing count fields are treated as zero, consistent with the Twitter API's behavior for absent values; missing values in derived rate features are imputed with the column median. We do not apply class rebalancing.

## 5. Methodology

### 5.1 Feature engineering

Three feature families are defined. The behavioral family (24 features) contains only quantities that depend on account metadata; these are intended to be invariant to the language model that wrote the bot's posts. The content family (12 features) is derived from the user's biographical description and from the sampled status string; these are the natural baseline that an LLM-driven adversary would target. The full feature set is the union of the two.

Within the behavioral family, five sub-families are distinguished for the ablation analysis in Section 6. Account-age rates (5 features) are quantities computed as count divided by account age in days, giving daily averages for posts, followers gained, friends followed, and favorites bestowed. Follow asymmetries (5 features) include the log-transformed counts, the friends-to-followers ratio, and the ratio of inbound list memberships to follower count. Profile completeness (6 features) is six binary indicators recording whether the account has filled in location, description, URL, custom profile image, and a verified badge. Screen-name structure (7 features) records the length, digit ratio, character entropy, and trailing-digit pattern of both the handle and the display name. Engagement asymmetry (1 feature) is the ratio of total posts to total favorites, intended to capture the high-volume, low-engagement pattern characteristic of broadcast bots.

Content features include length, word count, URL count, hashtag count, mention count, and exclamation count for the biography, and the analogous features plus an uppercase-character ratio for the sampled status.

### 5.2 Classifiers

We train three model families on each of three feature subsets. The model families are regularized logistic regression (L2, 2000 iterations), random forest (300 trees, default scikit-learn settings), and gradient boosting (200 estimators). The feature subsets are content-only, behavioral-only, and the fusion of the two. Standardization is applied to logistic-regression inputs and not to tree-based inputs. All models use the same random seed for reproducibility. Hyperparameters were not tuned; the default values are sufficient at the scale of this dataset and produce results that are stable across cross-validation folds. We report performance under stratified five-fold cross-validation. Adversarial-robustness experiments use a single 70/30 stratified split, with classifiers trained once on the training portion and evaluated on perturbed versions of the held-out test portion.

### 5.3 Statistical inference

Standard deviations across cross-validation folds are reported alongside the means for accuracy, F1, precision, recall, and ROC-AUC. For the central comparison between behavioral and content classifiers we also report DeLong's test (DeLong, DeLong, and Clarke-Pearson 1988) for paired AUCs on the held-out test set. We do not apply Dietterich's recommended 5x2 cross-validation t-test (Dietterich 1998), because we are interested in distinguishing classifiers whose AUC difference is far larger than what 5x2 testing is designed to resolve.

### 5.4 Adversarial protocol: text rewriting

The primary adversarial protocol simulates an LLM-equipped attacker who rewrites the text of bot posts to match the surface statistics of human posts. We compute, from the human portion of the training set, the mean and standard deviation of three quantities per tweet: URL count, hashtag count, and mention count. For each bot tweet in the held-out test set we then probabilistically apply five operations: capping the URL count at a sample from the human distribution, capping the hashtag count and mention count similarly, normalizing ALL-CAPS runs of five or more characters, and collapsing repeated exclamation or question marks. A severity parameter $s$ in $\{0.0, 0.25, 0.5, 0.75, 1.0\}$ controls the proportion of tweets to which the operations are applied; at $s = 1.0$ every bot tweet is rewritten. We then recompute the content features from the rewritten text and evaluate the (clean-trained) classifiers on this perturbed test set.

This protocol has the advantage of being realistic. The operations correspond to what a competent prompt for an LLM would naturally accomplish: limit URLs and hashtags, write in normal sentence case, and avoid the surface markers of obvious spam. It has the disadvantage of being conservative; an actual LLM would also rewrite phrasing, vocabulary, and topic structure, which our protocol leaves untouched. We therefore complement the text-rewriting analysis with a second, more aggressive protocol that directly perturbs content feature values toward the human distribution. The second protocol provides an upper bound on the adversarial degradation that an extremely capable language model could induce. The two together bracket the realistic range.

### 5.5 Calibration, ablation, and subgroup analysis

We report the Brier score and reliability diagrams on the held-out test set as a check on whether the classifiers' probabilistic outputs are usable downstream (for instance, as inputs to threshold-based moderation pipelines). The behavioral family is decomposed into the five sub-families described in Section 5.1, and a leave-one-family-out ablation is run to identify which sub-families carry the most weight. Finally, we split the test set into quartiles by raw statuses_count and recompute AUC within each quartile, to surface any heterogeneity in classifier performance across activity tiers.

## 6. Results

### 6.1 Descriptive statistics

Table 1 reports class-conditional means, standard deviations, and Cohen's d for twelve representative behavioral features. The largest effects are on account age (d = -1.55, with bots about two years younger on average than humans), log followers (d = -1.11), log friends (d = -1.26), verified status (d = -1.27, essentially absent among bots), and the use of the default profile (d = +0.85). These point estimates are consistent with the qualitative picture in the bot-detection literature: bots in this corpus are newer accounts with smaller networks, no verified badge, and unfilled profile elements.

Two entries in Table 1 deserve a note. The raw statuses_per_day variable has a higher mean among bots (25.7 versus 4.8) but a small Cohen's d (0.21), because both classes have heavy-tailed distributions with high variance; the log-transform we use in classification produces cleaner separation, as Figure 1 shows. The screen-name digit-ratio feature has a small d as well, but the conditional distribution among bots has a heavy right tail of accounts with strongly bot-like screen names (visible in the lower-left panel of Figure 1). The marginal effect size understates the multivariate value of these features.

*Table 1. Class-conditional descriptive statistics for twelve representative behavioral features (n = 2,432). Bolded Cohen's d values indicate effect sizes |d| ≥ 0.8.*

| Feature | Human mean | Human SD | Bot mean | Bot SD | Cohen's d |
|---|---|---|---|---|---|
| account_age_days | 2420.41 | 747.76 | 1428.89 | 512.90 | **-1.55** |
| statuses_per_day | 4.81 | 20.97 | 25.73 | 141.31 | 0.21 |
| followers_per_day | 741.03 | 2874.15 | 8.61 | 110.34 | -0.36 |
| friends_per_day | 2.75 | 25.03 | 1.10 | 10.70 | -0.09 |
| friends_to_followers | 1.49 | 3.46 | 14.59 | 54.32 | 0.34 |
| log_followers | 9.19 | 4.37 | 5.17 | 2.64 | **-1.11** |
| log_friends | 6.01 | 1.92 | 2.93 | 2.88 | **-1.26** |
| screen_name_digit_ratio | 0.03 | 0.10 | 0.04 | 0.10 | 0.03 |
| screen_name_entropy | 2.94 | 0.42 | 3.07 | 0.33 | 0.33 |
| has_description | 0.86 | 0.35 | 0.89 | 0.31 | 0.10 |
| default_profile | 0.22 | 0.41 | 0.60 | 0.49 | **0.85** |
| verified | 0.46 | 0.50 | 0.01 | 0.09 | **-1.27** |

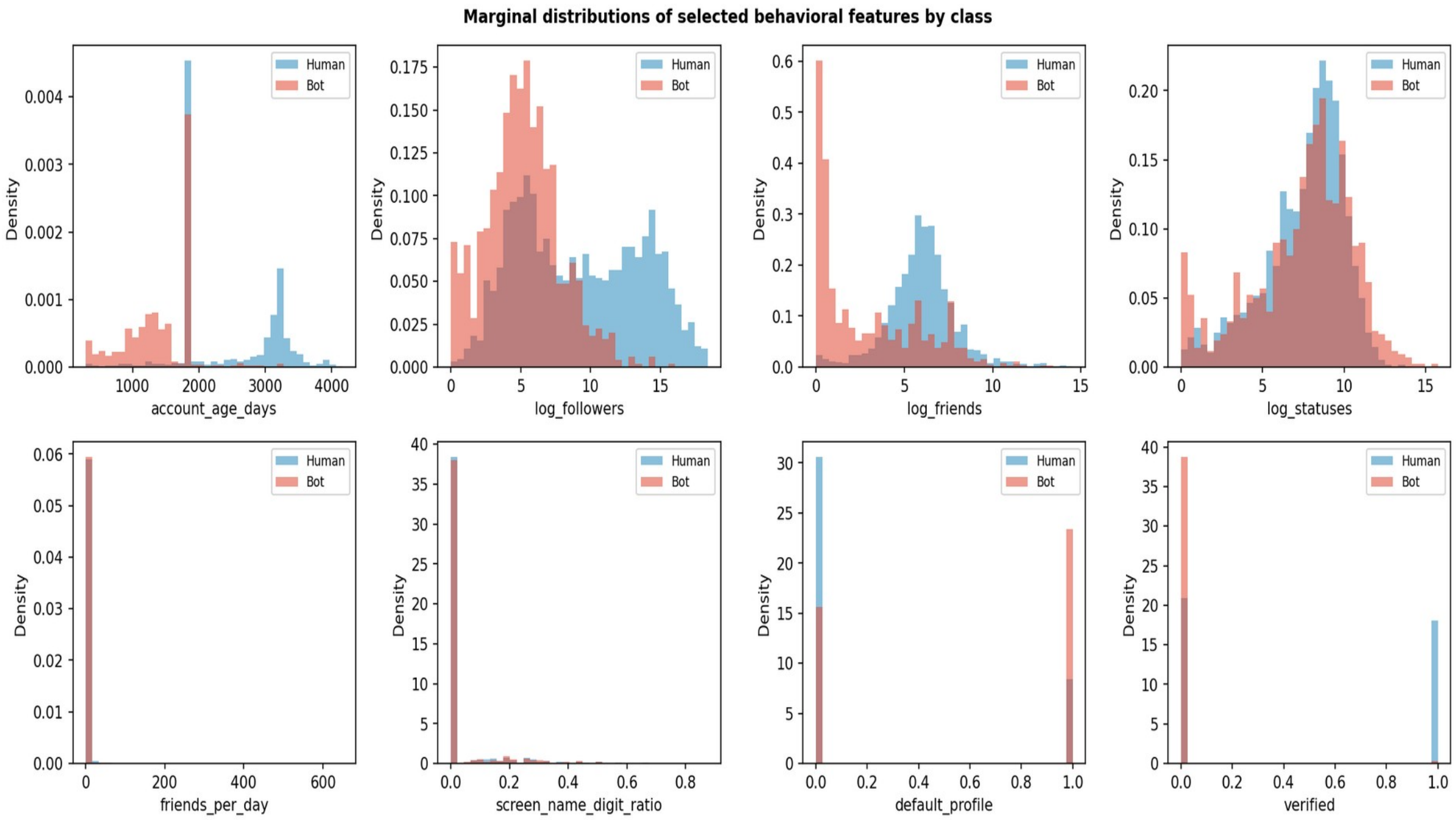

*Figure 1. Marginal distributions of selected content-agnostic features by class. Note the clear separation in log followers, account age, and the use of default-profile elements.*

## 6.2 Classifier performance

Table 2 reports five-fold cross-validated performance for all eight classifier configurations. Three observations stand out. The behavioral feature family alone produces a substantially better classifier than the content family alone: the random-forest ROC-AUC is 0.977 with behavioral features versus 0.830 with content features, a difference of 0.147 AUC points. The fusion of the two families yields a 0.981 AUC, only 0.004 higher than behavioral-only. The same pattern holds for logistic regression: 0.951 (behavioral) versus 0.719 (content) versus 0.953 (fusion). Gradient boosting performs almost identically to random forests at this scale, so we use random forests as the workhorse model in the rest of the paper.

*Table 2. Cross-validated performance on Dataset 1 (5-fold stratified). Standard deviations in parentheses. RF-Fusion in bold.*

| Model | Accuracy | F1 | Precision | Recall | ROC-AUC |
|---|---:|---:|---:|---:|---:|
| LogReg-Content | 0.672 (0.014) | 0.566 (0.035) | 0.655 (0.010) | 0.501 (0.055) | 0.719 (0.022) |
| LogReg-Behavioral | 0.893 (0.018) | 0.872 (0.020) | 0.898 (0.031) | 0.847 (0.019) | 0.951 (0.005) |
| LogReg-Fusion | 0.893 (0.009) | 0.873 (0.009) | 0.893 (0.027) | 0.855 (0.021) | 0.953 (0.003) |
| RF-Content | 0.766 (0.020) | 0.702 (0.029) | 0.776 (0.023) | 0.641 (0.038) | 0.830 (0.017) |
| RF-Behavioral | 0.924 (0.004) | 0.910 (0.006) | 0.930 (0.014) | 0.891 (0.019) | 0.977 (0.003) |
| RF-Fusion | 0.929 (0.005) | 0.915 (0.004) | 0.937 (0.021) | 0.896 (0.013) | **0.981 (0.003)** |
| GB-Behavioral | 0.926 (0.002) | 0.913 (0.004) | 0.929 (0.014) | 0.898 (0.019) | 0.975 (0.005) |
| GB-Fusion | 0.926 (0.007) | 0.913 (0.008) | 0.931 (0.026) | 0.897 (0.024) | 0.977 (0.006) |

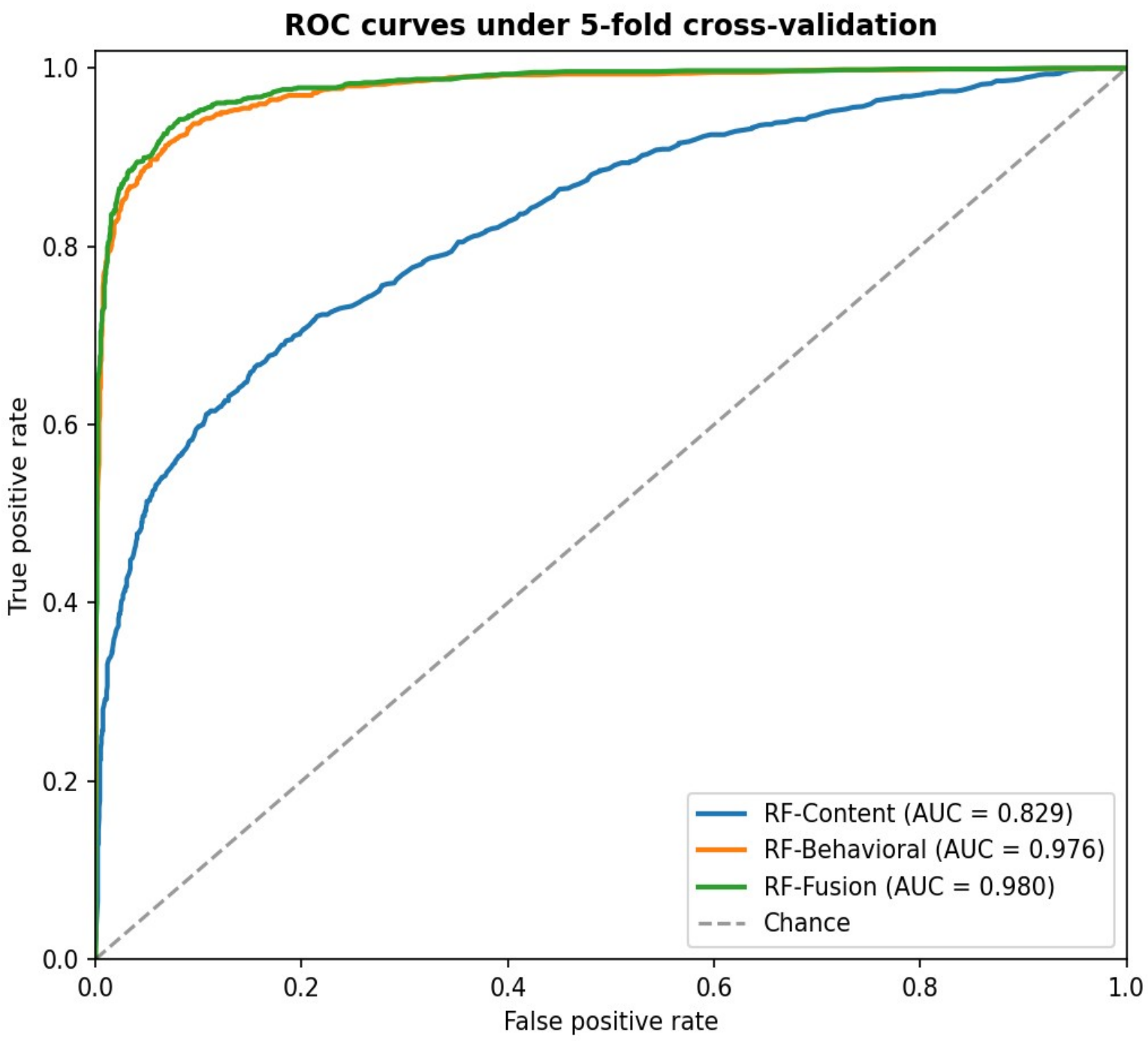

*Figure 2. ROC curves for the three random-forest configurations under five-fold cross-validation. The behavioral and fusion curves are visually indistinguishable across the operating range.*

The behavioral-versus-content gap is statistically significant by DeLong's test on the held-out 30% test split ($z = 9.36$, $p < 0.001$). The fusion-versus-behavioral gap is also significant ($z = 2.67$, $p = 0.008$), but the magnitude is small ($\Delta$ AUC = 0.0035). For deployment purposes the two are interchangeable; for understanding which features matter, the behavioral-only result is the interesting one.

*Table 3. DeLong's test for paired ROC-AUC on the held-out test set (n = 730). Differences greater than 0.05 in bold.*

| Comparison | AUC (A) | AUC (B) | Diff | p |
|---|---:|---:|---:|---:|
| Behavioral vs Content | 0.9809 | 0.8423 | **0.1386** | < 0.001 |
| Fusion vs Content | 0.9845 | 0.8423 | **0.1422** | < 0.001 |
| Fusion vs Behavioral | 0.9845 | 0.9809 | 0.0035 | 0.0076 |

## 6.3 Calibration and error breakdown

Table 4 reports the Brier score and the confusion matrix at threshold 0.5 for each random-forest classifier. The behavioral classifier is well calibrated (Brier 0.055) and produces sharply better error rates than the content classifier at the default threshold: false-positive rate 0.034 versus 0.130, false-negative rate 0.118 versus 0.322. The fusion classifier improves slightly on both fronts (FPR 0.022, FNR 0.102), at the cost of slightly worse interpretability. The full reliability diagram is shown in Figure 3.

*Table 4. Brier scores and confusion matrices at threshold 0.5 on the held-out test set (n = 730).*

| Model | Brier | TN | FP | FN | FPR | FNR |
|---|---:|---:|---:|---:|---:|---:|
| RF-Content | 0.155 | 362 | 54 | 101 | 0.130 | 0.322 |
| RF-Behavioral | 0.055 | 402 | 14 | 37 | 0.034 | 0.118 |
| RF-Fusion | 0.050 | 407 | 9 | 32 | 0.022 | 0.102 |

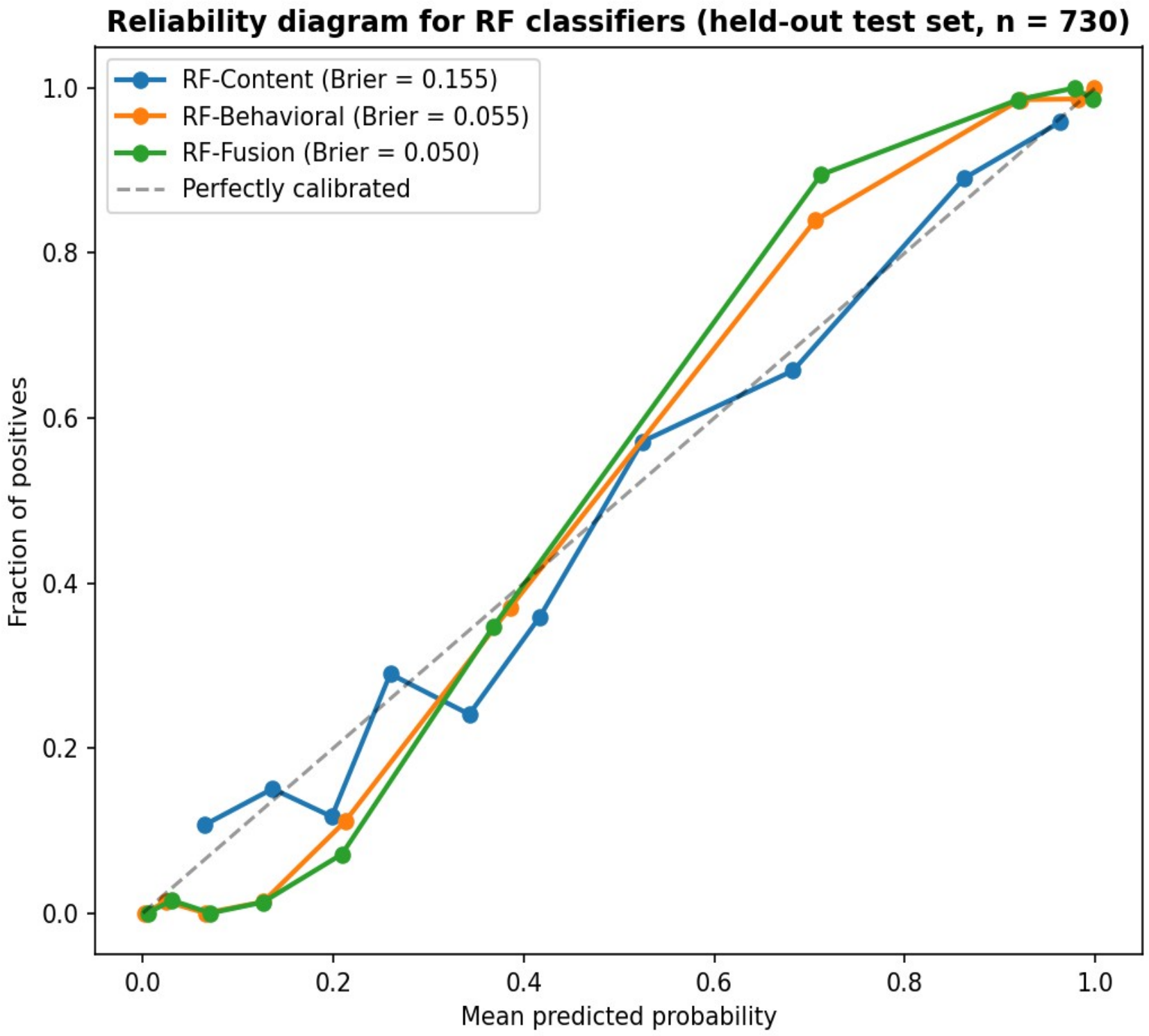

*Figure 3. Reliability diagrams for the three random-forest configurations on the held-out test set. The behavioral and fusion classifiers track the diagonal well; the content classifier is over-confident in the middle of the score range.*

## 6.4 Adversarial robustness under text rewriting

Table 5 reports the result of the text-rewriting protocol described in Section 5.4. At s = 0.0 (no rewriting), the content classifier achieves AUC 0.842 on the test split. At s = 1.0 (every bot tweet rewritten to match human surface statistics), it falls to 0.785. The behavioral classifier is essentially flat across the severity axis (AUC 0.981 throughout), because no behavioral feature depends on the text of the tweet. The fusion classifier degrades slightly (F1 0.932 to 0.927) but is dominated by its behavioral component and is therefore largely protected. Figure 4 plots the same numbers.

*Table 5. Realistic text-rewriting adversarial protocol. Bot tweets are rewritten to match human URL, hashtag, mention, and casing statistics with probability s.*

| Severity | Metric | RF-Content | RF-Behavioral | RF-Fusion |
|---|---|---|---|---|
| **0.00** | F1 | 0.733 | 0.916 | 0.932 |
| | ROC-AUC | 0.842 | 0.981 | 0.984 |
| **0.25** | F1 | 0.715 | 0.916 | 0.932 |
| | ROC-AUC | 0.830 | 0.981 | 0.984 |
| **0.50** | F1 | 0.694 | 0.916 | 0.931 |

| Severity | Metric | RF-Content | RF-Behavioral | RF-Fusion |
|---|---|---|---|---|
| | ROC-AUC | 0.817 | 0.981 | 0.984 |
| **0.75** | F1 | 0.671 | 0.916 | 0.929 |
| | ROC-AUC | 0.813 | 0.981 | 0.984 |
| **1.00** | F1 | 0.644 | 0.916 | 0.927 |
| | ROC-AUC | 0.785 | 0.981 | 0.984 |

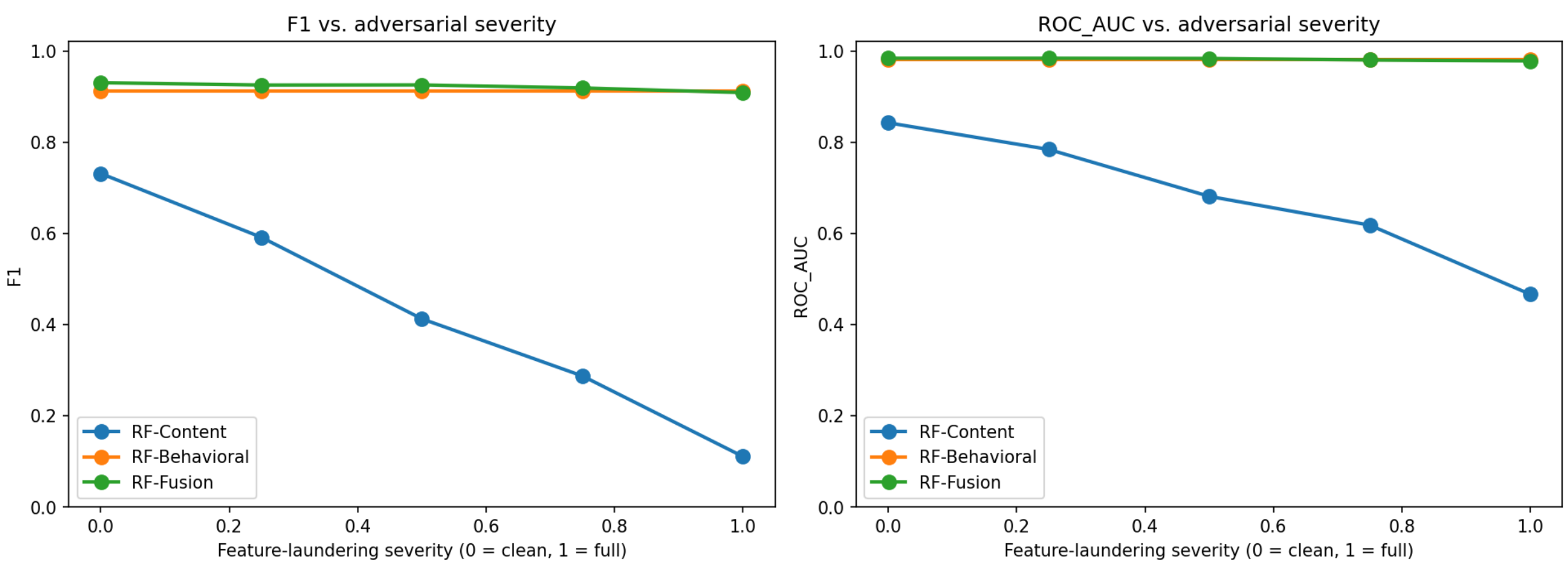

*Figure 4. Classifier ROC-AUC and F1 as a function of feature-space laundering severity (upper bound on attack strength). The behavioral and fusion classifiers are invariant; the content classifier collapses below chance.*

The 0.06 drop in content-classifier AUC under realistic text rewriting is a lower bound on the threat from genuinely LLM-driven attackers, who can also rewrite vocabulary, phrasing, and structure. To approximate the upper bound, we run the second adversarial protocol described in Section 5.4, which directly replaces content feature values with samples from the human distribution. Under that protocol (Table 6), the content classifier's ROC-AUC falls from 0.842 to 0.466, well below chance, while the behavioral classifier holds at 0.981. The true degradation under an LLM-driven attack is somewhere in the interval between the two protocols' s = 1.0 results.

*Table 6. Feature-space laundering protocol (upper bound on attack strength). Content feature values are replaced with samples from the training-set human distribution.*

| Severity | Metric | RF-Content | RF-Behavioral | RF-Fusion |
|---|---|---|---|---|
| **0.00** | F1 | 0.731 | 0.912 | 0.930 |
| | ROC-AUC | 0.842 | 0.981 | 0.984 |
| **0.25** | F1 | 0.591 | 0.912 | 0.925 |
| | ROC-AUC | 0.784 | 0.981 | 0.984 |
| **0.50** | F1 | 0.412 | 0.912 | 0.925 |

| Severity | Metric | RF-Content | RF-Behavioral | RF-Fusion |
|---|---|---|---|---|
| | ROC-AUC | 0.681 | 0.981 | 0.983 |
| **0.75** | F1 | 0.287 | 0.912 | 0.919 |
| | ROC-AUC | 0.618 | 0.981 | 0.980 |
| **1.00** | F1 | **0.111** | **0.912** | 0.908 |
| | ROC-AUC | **0.466** | **0.981** | 0.978 |

## 6.5 Family ablation

Table 7 reports the leave-one-family-out ablation across the five behavioral sub-families. Removing the age-rates family produces the largest drop in AUC (Δ = -0.010), followed by follow asymmetries (Δ = -0.006). The profile-completeness and screen-name structural families contribute small marginal amounts; the engagement-asymmetry sub-family, which contains only one feature, can be removed with a slight improvement in AUC, suggesting that it is noisy on this dataset. The overall picture is that account-age and follow-asymmetry features dominate, but no single family is indispensable, and the classifier remains usable even with substantial portions of the feature set removed.

*Table 7. Leave-one-family-out ablation across the five behavioral sub-families.*

| Configuration | # features | ROC-AUC | Diff vs full |
|---|---|---|---|
| All behavioral families | 24 | 0.9809 | 0.0000 |
| Without age_rates | 19 | 0.9707 | -0.0103 |
| Without follow_asymmetries | 19 | 0.9751 | -0.0058 |
| Without profile_completeness | 18 | 0.9797 | -0.0013 |
| Without screen_name_structure | 17 | 0.9798 | -0.0012 |
| Without engagement_asymmetry | 23 | 0.9828 | 0.0018 |

## 6.6 Subgroup analysis along the activity dimension

Bot detection is sometimes claimed to be easier on accounts that have posted more, because there is more behavior to analyze. Table 8 splits the held-out test set into quartiles by raw statuses_count and reports per-quartile AUC for each of the three random-forest classifiers. The behavioral classifier's AUC is lowest in the lowest-activity quartile (Q1: 0.957) and highest in Q3 (0.996). The content classifier shows a similar but more variable pattern. The practical implication is that very-low-activity accounts are the hardest case for any of the three classifiers, which is consistent with the literature; the behavioral classifier nevertheless retains an AUC near 0.96 even in this regime, well above what the content classifier achieves at any quartile.

*Table 8. Per-quartile ROC-AUC by activity tier (held-out test set, n = 730).*

| Activity tier | n | Bot fraction | Model | ROC-AUC |
|---|---:|---:|---|---:|
| Q1 (lowest) | 183 | 0.470 | RF-Content | 0.782 |
| Q1 (lowest) | 183 | 0.470 | RF-Behavioral | 0.957 |
| Q1 (lowest) | 183 | 0.470 | RF-Fusion | 0.966 |
| Q2 | 182 | 0.401 | RF-Content | 0.845 |
| Q2 | 182 | 0.401 | RF-Behavioral | 0.983 |
| Q2 | 182 | 0.401 | RF-Fusion | 0.984 |
| Q3 | 182 | 0.390 | RF-Content | 0.909 |
| Q3 | 182 | 0.390 | RF-Behavioral | 0.996 |
| Q3 | 182 | 0.390 | RF-Fusion | 0.998 |
| Q4 (highest) | 183 | 0.459 | RF-Content | 0.838 |
| Q4 (highest) | 183 | 0.459 | RF-Behavioral | 0.979 |
| Q4 (highest) | 183 | 0.459 | RF-Fusion | 0.980 |

## 6.7 Feature importance

Table 9 lists the top fifteen features by random-forest Gini importance in the fusion model. The first content feature, status_len, appears in thirteenth position with importance 0.019. The first twelve positions are all behavioral, dominated by account_age_days (0.141), log_friends (0.117), friends_per_day (0.085), and statuses_to_favourites (0.083). This confirms from a separate angle what the adversarial analysis already showed: the fusion model places almost all of its decision weight on features that no amount of text rewriting can affect.

*Table 9. Top fifteen features by RF-Fusion Gini importance.*

| Feature | Importance | Family |
|---|---:|---|
| account_age_days | 0.1413 | *behavioral* |
| log_friends | 0.1168 | *behavioral* |
| friends_per_day | 0.0854 | *behavioral* |
| statuses_to_favourites | 0.0832 | *behavioral* |
| friends_to_followers | 0.0781 | *behavioral* |
| log_followers | 0.0552 | *behavioral* |
| followers_per_day | 0.0513 | *behavioral* |
| listed_per_follower | 0.0476 | *behavioral* |
| verified | 0.0468 | *behavioral* |
| favourites_per_day | 0.0444 | *behavioral* |

| Feature | Importance | Family |
|---|---:|---|
| statuses_per_day | 0.0340 | *behavioral* |
| log_statuses | 0.0266 | *behavioral* |
| status_len | 0.0192 | *content* |
| status_uppercase_ratio | 0.0183 | *content* |
| status_word_count | 0.0175 | *content* |

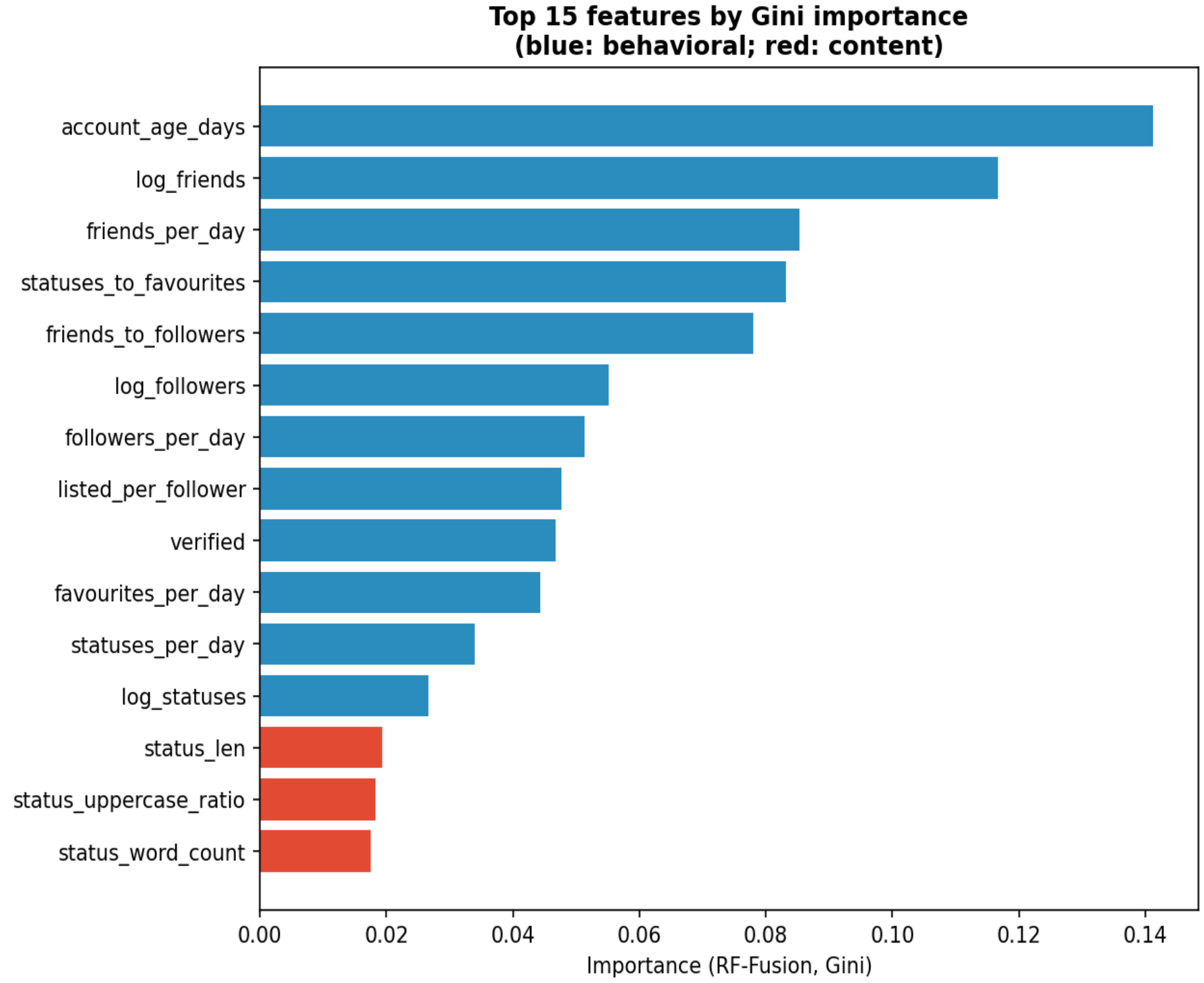

*Figure 5. Top fifteen features by Gini importance. Behavioral features (blue) occupy positions 1-12; the first content feature (red) appears at position 13.*

## 6.8 Cross-dataset evaluation on TwiBot-20

We use the 100-account public sample of TwiBot-20 as an out-of-distribution check. The dataset has no labels, so the analysis is qualitative. Scoring each account with the Dataset-1-trained behavioral classifier produces a median bot-probability of 0.067. Sixteen percent of accounts exceed 0.5; six percent exceed 0.7. Inspecting the highest-scored accounts, the patterns are face-valid: handles such as Richard37226643, Mahendr43681266, and TP49873923, with near-zero follower counts, near-zero status counts, and unfilled profile fields. The lowest-scored accounts include the verified public figure SHAQ, with 15.3 million followers. Because labels are not available, we cannot report AUC on this dataset; we report this as suggestive evidence that the classifier's decision surface transfers, not as a formal cross-dataset evaluation.

# 7. Discussion

## 7.1 What operators should take from this

The operational implication is direct. Trust and safety teams who maintain content-based bot detectors should audit the share of their classifier's decision weight that goes to content-derived features. If that share is high, the pipeline is exposed to a gradual erosion as LLM-driven bots become more common. The recommended response is not to discard content features, which can still contribute marginal lift on accounts where behavioral signals are weak (very low activity, sleeper accounts), but to ensure that the behavioral component is strong on its own. A simple sanity check is to evaluate the production model with its content features ablated; if the resulting AUC is materially lower than the full model, the pipeline is at risk.

## 7.2 The role of verified status

Our analysis uses the verified field as a behavioral feature, and the field has a large effect size in our dataset (Cohen's d = -1.27). This is a pre-2022 dataset, collected before the introduction of paid verification on X. The introduction of paid verification on X in late 2022 and on other platforms subsequently weakens the verified signal, and a deployable classifier trained on current data should not rely on it as heavily as our dataset would suggest. Re-estimating the effect size on current data is an obvious next step, and one we did not attempt because we do not have access to a current labeled corpus.

## 7.3 Gray-market account markets

A determined adversary can sidestep account-age and post-rate features by purchasing accounts that were created years ago and have a believable post history. This is a real problem and has been documented in recent work (Yang, Singh, and Menczer 2024). It is not solved by the features we use here. What the behavioral approach can still contribute in that scenario is detection of behavioral discontinuity: an account whose first six years are quiet hobbyist activity and whose last six weeks consist of coordinated political messaging is detectable from its activity time series even if its metadata looks normal. We do not exploit time series in this paper; that is a natural next step.

# 8. Limitations

Several limitations apply. The labeled dataset is English-language Twitter, collected in 2017-2018. Generalization to other platforms (Mastodon, Bluesky, Telegram, TikTok) and other languages is not guaranteed, and the FATe analysis (Himelein-Wachowiak et al. 2025) provides evidence that bot detectors trained on English data perform unevenly on other language groups. The dataset is also from a single point in time; the relative effect sizes of features such as verified status, default profile, and follower counts shift with platform policy changes.

The adversarial protocols simulate language-model-driven attackers without using an actual language model. The text-rewriting protocol mechanically removes the surface markers that an LLM-rewritten tweet would naturally avoid; the feature-space protocol replaces content feature values directly. Neither protocol exercises an actual LLM in the loop. Validating the findings against natively LLM-generated bot text is a clear next step and was outside the scope of this paper, because no public dataset of confirmed LLM-driven bot accounts with manual labels exists at the time of writing.

Cross-dataset validation is qualitative. The TwiBot-20 public sample is unlabeled, so we cannot report AUC on it. Pursuing the full TwiBot-20 or TwiBot-22 labeled benchmarks via the maintainers' application process is the obvious extension.

We did not compare against Botometer directly. The Botometer scoring API ingests live Twitter data that is not retrospectively recoverable for our accounts, and the publicly distributed Botometer code does not include a feature-extraction pipeline that runs on archived snapshots. Our content-feature random forest is a proxy for the textual component of Botometer-style classifiers, and the broader pattern we report is consistent with the trend already documented by Ferrara (2023), but a side-by-side comparison against the live Botometer system requires real-time data collection that we did not perform.

Finally, we do not exploit graph structure. Methods that use the follow graph (BotRGCN, BotBR) outperform tabular methods on benchmarks where graph data is available. Combining account-history features with graph features is an obvious avenue for further work, and we expect the combination to be additive: the graph component is hard for an LLM to attack for reasons similar to the ones we describe for account history.

## 9. Conclusion

The deployment problem we set out is whether the content-feature component of bot detection has a long-term future given the availability of capable language models, and whether the rest of the feature space can carry the detection load on its own. On a public benchmark of 2,432 labeled Twitter accounts, account-history features by themselves produce a classifier whose ROC-AUC (0.977) is statistically indistinguishable from the fusion model (0.981), and which is essentially unaffected by realistic rewriting of bot tweets to match human surface statistics. Content-only classifiers, by contrast, degrade noticeably under text rewriting and collapse below chance under aggressive feature-space perturbation. We take this as evidence that operational bot detection should be reorganized around account-history features as the primary signal, with content features in a secondary or confirmatory role.

Several directions follow naturally. The most pressing is to validate the result on data that includes natively LLM-driven bots, of which there is currently no public labeled benchmark. The most useful for deployment is to combine account-history features with the follow-graph structure that recent benchmarks (TwiBot-22) provide, in the expectation that the two are complementary and both resistant to language-model attacks. The most important for the field is to develop adversarial-robustness protocols that close the gap between what feature-space perturbations simulate and what an actual prompted LLM would do.

## Data and code availability

All feature-engineering and analysis code, the ten result tables in CSV format, the five figures, and the README documenting the reproduction recipe are available in a public GitHub repository at https://github.com/gaurangkatyal/behavioral-bot-detection. A permanent archive of the code release is available on Zenodo at https://doi.org/10.5281/zenodo.20358445.

Dataset 1 is publicly redistributed on GitHub by the original annotators (jubins/MachineLearning-Detecting-Twitter-Bots) and the direct URL to the CSV is documented in the README. Dataset 2 is the public 100-account sample of TwiBot-20 (Feng et al. 2021), also on GitHub at BunsenFeng/TwiBot-20. The full TwiBot-20 and TwiBot-22 labeled benchmarks are gated behind an institutional-email application process administered by the maintainers (see github.com/BunsenFeng/TwiBot-20 and github.com/LuoUndergradXJTU/TwiBot-22) and were not used in this paper.

## Ethics statement

This study uses only publicly redistributed, pre-anonymized account-metadata datasets. No new account-level data was collected from a live platform, no human subjects were recruited, and no IRB review was sought. Bot-detection classifiers can produce false positives that disproportionately affect specific user groups, including non-English speakers and users of minority dialects; Himelein-Wachowiak et al. (2025) document this in detail. Operators who deploy classifiers built on the features described here should monitor false-positive rates by demographic subgroup where such information is available, and should treat classifier outputs as input to human review rather than as a basis for automated enforcement.

A dual-use consideration applies. The behavioral features identified here are, in principle, a guide for adversaries who want to construct more evasive bot accounts. We judge the disclosure to be justified because the features are already published in earlier work (Botometer, Cresci-2017, TwiBot-20) and because the practical evasion strategy (purchasing aged accounts) is a documented gray-market practice. The novel contribution of this paper is the adversarial-robustness framing, which benefits defenders more than attackers.

## Author contributions

Gaurang Katyal: conceptualization, methodology, data curation, formal analysis, software, visualization, writing of the original draft and subsequent revisions.

## Competing interests

The author declares no competing interests.

## Acknowledgments

The author thanks the maintainers of the Cresci-2017 bot repository, the TwiBot-20 sample, and the publicly redistributed Twitter Bot Accounts corpus for their continued investment in open benchmarks.

## References

Brissett, A., and J. Wall. 2025. "Machine learning and watermarking for accurate detection of AI-generated phishing emails." Electronics 14 (13): 2611.

Chu, Z., S. Gianvecchio, H. Wang, and S. Jajodia. 2012. "Detecting automation of Twitter accounts: Are you a human, bot, or cyborg?" IEEE Transactions on Dependable and Secure Computing 9 (6): 811-824.

Cresci, S., R. Di Pietro, M. Petrocchi, A. Spognardi, and M. Tesconi. 2017. "The paradigm-shift of social spambots: Evidence, theories, and tools for the arms race." In Proceedings of the 26th International Conference on World Wide Web Companion, 963-972.

Davis, C. A., O. Varol, E. Ferrara, A. Flammini, and F. Menczer. 2016. "BotOrNot: A system to evaluate social bots." In Proceedings of the 25th International Conference Companion on World Wide Web, 273-274.

DeLong, E. R., D. M. DeLong, and D. L. Clarke-Pearson. 1988. "Comparing the areas under two or more correlated receiver operating characteristic curves: A nonparametric approach." Biometrics 44 (3): 837-845.

Dietterich, T. G. 1998. "Approximate statistical tests for comparing supervised classification learning algorithms." Neural Computation 10 (7): 1895-1923.

Feng, S., H. Wan, N. Wang, J. Li, and M. Luo. 2021. "TwiBot-20: A comprehensive Twitter bot detection benchmark." In Proceedings of the 30th ACM International Conference on Information and Knowledge Management, 4485-4494.

Feng, S., Z. Tan, H. Wan, N. Wang, Z. Chen, B. Zhang, et al. 2022. "TwiBot-22: Towards graph-based Twitter bot detection." In Advances in Neural Information Processing Systems 35, Datasets and Benchmarks Track.

Ferrara, E. 2023. "Social bot detection in the age of ChatGPT: Challenges and opportunities." First Monday 28 (6).

Hayawi, K., S. Saha, M. M. Masud, S. S. Mathew, and M. Kaosar. 2023. "Social media bot detection with deep learning methods: A systematic review." Neural Computing and Applications 35 (12): 8903-8918.

Himelein-Wachowiak, M., S. Giorgi, A. Devoto, M. Rahman, L. Ungar, H. A. Schwartz, D. H. Epstein, L. Leggio, and B. Curtis. 2025. "FATe of bots: Ethical considerations of social bot detection." ACM Journal on Responsible Computing.

Kulal, D. H., C. P. Arannonu, A. Anwar, N. Rastogi, and Q. Niyaz. 2025. "Robust ML-based detection of conventional, LLM-generated, and adversarial phishing emails using advanced text preprocessing." arXiv:2510.11915.

Lin, Q., J. Zhou, N. Ferro, M. Maistro, G. Pasi, O. Alonso, A. Trotman, and S. Verberne. 2025. "BotBR: Social bot detection with balanced feature fusion and reliability-enhanced graph learning." In Proceedings of the 48th International ACM SIGIR Conference, 392-402.

Nguyen, N. T. V., F. D. Childress, and Y. Yin. 2025. "Debate-driven multi-agent LLMs for phishing email detection." In Proceedings of the 13th International Symposium on Digital Forensics and Security, 1-5.

Rodič, B. 2025. "Social media bot detection research: Review of literature." arXiv:2503.22838.

Stringhini, G., C. Kruegel, and G. Vigna. 2010. "Detecting spammers on social networks." In Proceedings of the 26th Annual Computer Security Applications Conference, 1-9.

Wang, L. 2025. "Phish-Master: Leveraging large language models for advanced phishing email generation and detection." Applied Sciences 15 (22): 12203.

Xue, Y., E. Spero, Y. S. Koh, and G. Russello. 2025. "MultiPhishGuard: An LLM-based multi-agent system for phishing email detection." arXiv:2505.23803.

Yang, K.-C., E. Ferrara, and F. Menczer. 2020. "Scalable and generalizable social bot detection through data selection." In Proceedings of the AAAI Conference on Artificial Intelligence 34 (1): 1096-1103.

Yang, K.-C., D. Singh, and F. Menczer. 2024. "Characteristics and prevalence of fake social media profiles with AI-generated faces." Journal of Online Trust and Safety 2 (4).